\documentclass[prd,nofootinbib,showpacs,preprint]{revtex4}
\usepackage{amsmath}
\usepackage{graphicx}
\usepackage{hyperref}  
\usepackage{color} 
\usepackage{epsfig}
\usepackage{epsf}

\newcommand{\vev}[1]{\langle #1 \rangle}

\begin{document}
\title{Spontaneous R-parity breaking, Left-Right Symmetry and Consistent Cosmology with Transitory Domain Walls \\}
\author{Debasish Borah}
\email{debasish@phy.iitb.ac.in}
\affiliation{Department of Physics, Indian Institute of Technology, 
Bombay, Mumbai - 400076, India}
\author{Sasmita Mishra}
\email{sasmita@phy.iitb.ac.in}
\affiliation{Department of Physics, Indian Institute of Technology, 
Bombay, Mumbai - 400076, India}

\begin{abstract}
Domain wall formation is quite generic in spontaneous Left-Right parity (D-parity) 
breaking models. Since they are in conflict with cosmology, we need 
some mechanisms to remove them. Planck scale suppressed effects have been 
considered to be quite successful for this purpose. We study this possibility in Minimal Supersymmetric Left-Right (SUSYLR) model originally proposed by Kuchimanchi et al \cite{Kuchimanchi:1993jg} where both D-parity and R-parity $(R_p = (-1)^{3(B-L)+2s})$  are spontaneously broken. We find that Planck scale suppressed terms allowed for the specific particle content of this model can successfully remove the domain walls provided the D-parity breaking scale is relatively low $(\leq 10^5-10^7 \text{GeV})$. However, demanding this theory to be part of a grand unified theory such as $SO(10)$ forces the D-parity breaking scale to be very high $(\geq 10^{14} \text{GeV})$ and hence is in conflict with the constraint from domain wall removal. We also find another class of R-parity violating SUSYLR models where both these constraints can be simultaneously satisfied.
\end{abstract}

\maketitle

\section{Introduction}
Left-Right Symmetric Models (LRSM) \cite{Pati:1974yy,Mohapatra:1974gc, Senjanovic:1975rk, 
Mohapatra:1980qe, Deshpande:1990ip} provide a framework within which spontaneous parity breaking as well as tiny neutrino masses \cite{Fukuda:2001nk, Ahmad:2002jz, 
Ahmad:2002ka, Bahcall:2004mz} can be successfully implemented without reference to very high scale physics such as grand unification. Incorporating
Supersymmetry (SUSY) into it comes with other advantages like
providing a solution to the gauge hierarchy problem, and providing 
a Cold Dark Matter candidate which is the lightest
supersymmetric particle (LSP). In Minimal Supersymmetric
Standard Model (MSSM), the stability of LSP is guaranteed by R-parity,
defined as $R_p =(-1)^{3(B-L)+2S}$ where $S$ is the spin of the particle.
This is a discrete symmetry put by hand in MSSM to keep the baryon
number (B) and lepton number (L) violating terms away from the
superpotential. In generic implementations of Left-Right
symmetry, R-parity is a part of the gauge symmetry and hence not ad-hoc like in
the MSSM. In one class of models ~\cite{Aulakh:1998nn, Aulakh:1997ba,Babu:2008ep,Patra:2009wc}, spontaneous parity breaking is 
achieved without breaking R-parity. This was not possible in minimal supersymmetric left right(SUSYLR) models where the only way to break parity is to consider 
spontaneous R-parity violation \cite{Kuchimanchi:1993jg}. In minimal
SUSYLR model parity, $SU(2)_R$ gauge symmetry as well as R-parity break 
simultaneously by the vacuum expectation value of right handed sneutrino. 

Spontaneous breaking of exact discrete symmetries like parity (which we shall denote as
D-parity hereafter), as well as R-parity have got cosmological
implications since they lead to frustrated phase transitions leaving behind a network of domain walls (DW). These 
domain walls, if not removed will be in conflict with the observed Universe \cite{Kibble:1980mv,Hindmarsh:1994re}. It was pointed 
out \cite{Rai:1992xw,Lew:1993yt} that Planck scale suppressed non-renormalizable operators 
can be a source of domain wall instability. Supersymmetry dictates the structure of these non-renormalizable 
terms but also gives rise to the gravitino overabundance problem. Incorporating all these restrictions, the constraint on 
the D-parity breaking scale in R-parity conserving SUSYLR models 
~\cite{Aulakh:1998nn, Aulakh:1997ba,Babu:2008ep} 
has been discussed in \cite{Mishra:2009mk}. Here we extend the analysis to a more general 
class of models where both R-parity and D-parity break spontaneously. It should be mentioned that the formation of domain walls is 
not generic in all Left-Right models. Models where D-parity and $SU(2)_R$ gauge symmetry are broken at two different stages do not suffer from this problem \cite{Chang:1983fu,Chang:1984uy,Sarkar:2007er,Patra:2009wc,Dev:2009aw,Borah:2010zq,Borah:2011zz}. In these models, the vacuum expectation value (vev) of a parity odd singlet field breaks the D-parity first and $SU(2)_R$ gauge symmetry gets broken at a later stage by either Higgs triplets and Higgs doublets.

This paper is organized as follows. In section \ref{sec:DW-section} we briefly review the 
domain wall dynamics. In section \ref{sec:SUSYLR1} we discuss minimal SUSYLR model with
Higgs triplets, constraints on the symmetry breaking scale from successful removal of domain wall as well as gauge coupling unification and in section \ref{sec:SUSYLR2} we do this 
same analysis for minimal SUSYLR model with Higgs doublets. We summarize our results in 
section \ref{sec:con}.

\section{Domain Wall dynamics}
\label{sec:DW-section}

Discrete symmetries and their spontaneous breaking are both common instances and desirable in 
model building. The spontaneous breaking of such discrete symmetries gives rise to a network of 
domain walls leaving the accompanying phase transition frustrated \cite{Kibble:1980mv, Hindmarsh:1994re}. 
The danger of a frustrated phase transition can therefore be evaded if a small explicit breaking of discrete 
symmetry can be introduced.

Due to the smallness of such discrete symmetry breaking, the resulting domain walls may be relatively long lived and
 can dominate the Universe for a long time. Since this will be in conflict with the observed Universe, these domain 
walls need to disappear at a very high energy scale (at least before Big Bang Nucleosynthesis). Keeping this in mind, we summarize the three cases 
of domain wall dynamics discussed in \cite{Mishra:2009mk}, one of which originates in radiation dominated (RD) Universe and destabilized also within
the radiation dominated Universe. This scenario was originally proposed by Kibble \cite{Kibble:1980mv} and Vilenkin \cite{Vilenkin:1984ib}. The second scenario was
essentially proposed in \cite{Kawasaki:2004rx}, which consists of the walls
originating  in a radiation dominated phase, subsequent to which the 
Universe enters a matter dominated (MD) phase, either due to substantial 
production of heavy unwanted relics such as moduli, or simply due
to a coherent oscillating scalar field. The third one is a variant of the 
MD model in which the domain walls dominate the Universe for a considerable epoch giving rise to a mild inflationary behavior or weak inflation (WI) \cite{Lyth:1995hj,Lyth:1995ka}. In all these cases the domain
walls disappear before they come to dominate the energy density of the Universe.

When a scalar field $\phi$ acquires a vev at a scale $M_R$ at some critical temperature $T_c$, a phase transition occurs leading to the formation of domain walls. The energy density trapped per unit area of such a wall is $\sigma \sim  M^3_R$. The dynamics of the walls are determined by two quantitites, force due to tension $f_T \sim \sigma/R$ and force due to friction $f_F \sim \beta T^4$ where $R$ is the average scale of radius of curvature prevailing in the wall complex, $\beta$ is the speed at which the domain wall is navigating through the medium and $T$ is the temperature. The epoch at which these two forces balance each other sets the time scale $t_R \sim R/\beta $. Putting all these together leads to the scaling law for the growth of the scale $R(t)$:
\begin{equation}
R(t) \approx (G \sigma)^{1/2}t^{3/2}
\end{equation}
The energy density of the domain walls goes as $\rho_W \sim (\sigma R^2/R^3) \sim (\sigma/Gt^3)^{1/2} $. In a radiation dominated era this $\rho_W$ is comparable to the energy density of the Universe $[\rho \sim 1/(Gt^2)]$ around time $t_0 \sim 1/(G \sigma)$. 

The pressure difference arising from small asymmetry on the two sides of the wall competes with the two forces $f_F \sim 1/(Gt^2)$ and $f_T \sim (\sigma/(Gt^3))^{1/2}$ discussed above. For $\delta \rho$ to exceed either of these two quantities before $t_0 \sim 1/(G\sigma)$
\begin{equation}
 \delta \rho \geq G\sigma^2 \approx \frac{M^6_R}{M^2_{Pl}} \sim M_R^4 \left(\frac{M_R}{M_{Pl}}\right)^2
\label{eq:RD-delta-rho}
\end{equation}

Similar analysis in the matter dominated era, originally considered in \cite{Kawasaki:2004rx} begins with the assumption that the initially formed wall complex in a phase transition is expected to rapidly relax to a few walls per horizon volume at an epoch characterized by Hubble parameter value $H_i$. Thus the initial energy density of the wall complex is $\rho^{\text{in}}_W \sim \sigma H_i$. This epoch onward the energy density of the Universe is assumed to be dominated by heavy relics or an oscillating modulus field and in both the cases the scale factor grows as $a(t) \propto t^{2/3}$. The energy density scales as $\rho_{\text{mod}} \sim \rho^{\text{in}}_{\text{mod}}/(a(t))^3$. If the domain wall (DW) complex remains frustrated, i.e. its energy density contribution $\rho_{\text{DW}} \propto 1/a(t)$, the Hubble parameter at the epoch of equality of DW contribution with that of the rest of the matter is given by \cite{Kawasaki:2004rx}
\begin{equation}
H_{\text{eq}} \sim \sigma^{3/4}H^{1/4}_i M^{-3/2}_{Pl}
\label{heq}
\end{equation}
Assuming that the domain walls start decaying as soon as they dominate the energy density of the Universe, which corresponds to a temperature $T_D$ such that $H^2_{\text{eq}} \sim GT^4_D$, the above equation gives 
\begin{equation}
T^4_D \sim \sigma^{3/2} H^{1/2}_i M^{-1}_{Pl}
\label{td}
\end{equation}
Under the assumption that the domain walls are formed at $T \sim \sigma^{1/3}$
\begin{equation}
H^2_i = \frac{8\pi}{3}G\sigma^{4/3} \sim \frac{\sigma^{4/3}}{M^2_{Pl}}
\end{equation}
Now from Eq. (\ref{td})
\begin{equation}
T^4_D \sim \frac{\sigma^{11/6}}{M^{3/2}_{Pl}} \sim \frac{M^{11/2}_R}{M^{3/2}_{Pl}} \sim M^4_R \left(\frac{M_R}{M_{Pl}}\right)^{3/2}
\end{equation}
Demanding $\delta \rho > T^4_D$ leads to
\begin{equation}
\delta \rho >  M^4_R \left(\frac{M_R}{M_{Pl}}\right)^{3/2}
\end{equation}

The third possibility is the walls dominating the energy density of the Universe for a limited epoch which leads to a mild inflation. This possibility was considered in \cite{Lyth:1995hj,Lyth:1995ka}. As discussed in \cite{Mishra:2009mk}, the evolution of energy density of such walls can be expressed as 
\begin{equation}
\rho_{\text{DW}}(t_d) \sim \rho_{\text{DW}}(t_{\text{eq}}) (\frac{a_{\text{eq}}}{a_d}) 
\end{equation}
where $a_{\text{eq}} (a_d)$ is the scale factor at which domain walls start dominating (decaying) and $t_{\text{eq}} (t_d)$ is the corresponding time. If the epoch of domain wall decay is characterized by temperature $T_D$, then $\rho_{\text{DW}} \sim T^4_D$ and the above equation gives 
\begin{equation}
T^4_D = \rho_{\text{DW}} (t_{\text{eq}}) (\frac{a_{\text{eq}}}{a_d}) 
\label{td1}
\end{equation}
In the matter dominated era the energy densit of the moduli fields scale as 
\begin{equation}
\rho^d_{\text{mod}} \sim \rho^{\text{eq}}_{\text{mod}} (\frac{a_{\text{eq}}}{a_d})^3
\end{equation}
Using this in equation (\ref{td1}) gives 
\begin{equation}
\rho^d_{\text{mod}} \sim \frac{T^{12}_D}{\rho^2_{\text{DW}} (t_{\text{eq}})}
\end{equation}
Domain walls start dominating the Universe after the time of equality,  $\rho_{\text{DW}}(t_d) > \rho^d_{\text{mod}} $. So the pressure difference across the walls when they start decaying is given by 
\begin{equation}
\delta \rho \geq \frac{T^{12}_D G^2}{H^4_{\text{eq}}}
\end{equation}
where $H^2_{\text{eq}} \sim G \rho_{\text{DW}} (t_{\text{eq}})$. Replacing the value of $H_{\text{eq}}$ from equation (\ref{heq}), the pressure difference becomes 
\begin{equation}
\delta \rho \geq M^4_R \frac{T^{12}_D M^3_{Pl}}{M^{15}_R}
\end{equation}
Unlike the previous two cases RD and MD, here it will not be possible to estimate $T_D$ in terms of other mass scales and we will keep it as undetermined.

\section{Minimal Supersymmetric Left-Right Model (MSLRM) with Higgs triplets}
\label{sec:SUSYLR1}
We consider the minimal SUSYLR model of Kuchimanchi et al \cite{Kuchimanchi:1993jg}
in this section. Although the minimality of the Higgs content is an attractive feature of this model, the authours concluded that D-parity can be spontaneously broken only at the expense of breaking R-parity spontaneously at the same energy scale by the vev of right handed sneutrino. Since the R-parity violation(RPV) is in the leptonic sector only, the dangerous proton decay problem can be evaded in this model. 
The models where left handed sneutrino vev gives rise to RPV are strongly disfavored by electroweak precision measurements \cite{GonzalezGarcia:1989zh,Romao:1989yh}. However there is no such strict constraints on models where right handed sneutrino vev gives rise to RPV provided the extra gauge boson masses lie above the allowed lower bounds \cite{Amsler:2008zzb,Beall:1981ze}. Here we find another constraint on this RPV scale from domain wall removal as well as gauge coupling unification. 

The matter content of this model is 
\begin{equation}
Q=
\left(\begin{array}{c}
\ u \\
\ d
\end{array}\right)
\sim (3,2,1,\frac{1}{3}),\hspace*{0.8cm}
Q_c=
\left(\begin{array}{c}
\ d_c \\
\ u_c
\end{array}\right)
\sim (3^*,1,2,-\frac{1}{3}),\nonumber 
\end{equation}
\begin{equation}
L=
\left(\begin{array}{c}
\ \nu \\
\ e
\end{array}\right)
\sim (1,2,1,-1), \quad
L_c=
\left(\begin{array}{c}
\ \nu_c \\
\ e_c
\end{array}\right)
\sim (1,1,2,1)
\end{equation}
The Higgs sector of this minimal consists  of the Higgs bidoublets and Higgs triplets
\begin{equation}
\Phi_1=
\left(\begin{array}{cc}
\ \phi^0_{11} & \phi^+_{11} \\
\ \phi^-_{12} & \phi^0_{12}
\end{array}\right)
\sim (1,2,2,0),\hspace*{0.2cm} 
\Phi_2=
\left(\begin{array}{cc}
\ \phi^0_{21} & \phi^+_{21} \\
\ \phi^-_{22} & \phi^0_{22}
\end{array}\right)
\sim (1,2,2,0), \nonumber 
\end{equation}
\begin{equation}
\Delta =
\left(\begin{array}{cc}
\ \delta^+_L/\surd 2 & \delta^{++}_L \\
\ \delta^0_L & -\delta^+_L/\surd 2
\end{array}\right)
\sim (1,3,1,2), \hspace*{0.2cm}
\bar{\Delta} =
\left(\begin{array}{cc}
\ \Delta^-_L\surd 2 & \Delta^0_L \\
\ \Delta^{--}_L & -\Delta^-_L/\surd 2
\end{array}\right)
\sim (1,3,1,-2),\nonumber 
\end{equation}
\begin{equation}
\Delta_c =
\left(\begin{array}{cc}
\ \delta^+_R/\surd 2 & \delta^{++}_R \\
\ \delta^0_R & -\delta^+_R/\surd 2
\end{array}\right)
\sim (1,1,3,-2), \hspace*{0.2cm}
\bar{\Delta}_c =
\left(\begin{array}{cc}
\ \Delta^-_R/\surd 2 & \Delta^0_R \\
\ \Delta^{--}_R & -\Delta^-_R/\surd 2
\end{array}\right)
\sim (1,1,3,2) \nonumber
\end{equation}
The renormalizable superpotential is 
\begin{eqnarray}
W_{ren} &=& h^{(i)}_l L^T\tau_2 \Phi_i \tau_2 L_c+ h^{(i)}_q Q^T\tau_2 
\Phi_i \tau_2 Q_c+  i f L^T\tau_2 \Delta L +i f L^T_c 
\tau_2 \Delta_c L_c \\ \nonumber
&+& m_{\Delta} \text{Tr}\Delta \bar{\Delta}
+m_{\Delta} \text{Tr}\Delta_c \bar{\Delta}_c 
+\mu_{ij}\text{Tr}\tau_2\Phi^T_i\tau_2\Phi_j 
\end{eqnarray}
where $h^{(i)}_{q,l}=h^{(i)\dagger}_{q,l},\mu_{ij}=\mu{ji}=
\mu^*_{ij}$ and $f,h$ are symmetric matrices.
 It has been shown that with this minimal field content
it is not possible to break the D-parity spontaneously.
Adding a parity odd singlet also does not improve the situation.
The authors showed that the D-parity breaking vacua in this case also give rise to the breaking of electromagnetic charge.

The authors \cite{Kuchimanchi:1993jg} proposed an alternative scenario where it was shown
that by allowing a non-zero vev for right handed sneutrino, $\tilde{\nu}_c$
it is possible to get D-parity breaking minima which preserve electromagnetic charge. However, the vev of sneutrino which has odd $U(1)_{B-L}$ charge also gives rise to spontaneous R-parity violation. Here we follow the approximations adopted by the authors to find a region in parameter space of the coupling constants giving rise to
the desired minima. The first approximation is the one where they choose
the parameter space, such that $g^2$ and $g'^2$ are smaller than 
the constants $h^2$ and $f^2$. With this approximation the $D$-terms
become weaker than the trilinear terms that contain the triplet
scalars and the sleptons. The second approximation is made in order
to maintain the hierarchy between the electroweak scale and the
parity breaking scale, i.e. $\vev{L_c},\,\vev{\Delta_c} \gg  \vev{\Phi}$.
With these approximations the scalar potential can now be written as,
\begin{eqnarray}
V &=& m_l^2(\tilde{L}^{\dagger}\tilde{L}+\tilde{L}^{\dagger}_c\tilde{L}_c)
+M^2_1\text{Tr}(\Delta \Delta^{\dagger}+\Delta_c \Delta^{\dagger}_c)+
M^2_2 \text{Tr}(\bar{\Delta}
 \bar{\Delta}^{\dagger}+ \bar{\Delta}_c \bar{\Delta}^{\dagger}_c)+
\lvert h \rvert^2\tilde{L}^{\dagger}_c \tilde{L}_c \tilde{L}^{\dagger}\tilde{L}\nonumber \\
&& +\lvert f \rvert^2 [(\tilde{L}^{\dagger}\tilde{L})^2+ 
(\tilde{L}^{\dagger}_c \tilde{L}_c)^2 ]+4 \lvert f 
\rvert^2(\lvert \tilde{L}^T_c\tau_2 \Delta_c \rvert^2+\lvert 
\tilde{L}^T\tau_2 \Delta \rvert^2)+M'^2 \text{Tr}
(\Delta \bar{\Delta}+\Delta_c \bar{\Delta}_c +h.c.) \nonumber \\
&& +[ \tilde{L}^T\tau_2 (iv\Delta+iM^*f\bar{\Delta}^{\dagger})
\tilde{L}+\tilde{L}^T_c\tau_2 (iv\Delta_c+iM^*F\bar{\Delta}^{\dagger}_c)
\tilde{L}_c +h.c. ]
\label{eq:Scalar-potential}
\end{eqnarray}
Consider the case where the Right handed fields getting
a non-zero vev, at the same time the vev for the Left handed
fields is zero. So,
$$ \langle L \rangle = 0, \quad \langle \Delta \rangle =
 \langle \bar{\Delta} \rangle = 0 $$
$$ \langle L_c \rangle = \left(\begin{array}{c}
\ l_c \\
\ 0
\end{array}\right), \quad \langle \Delta_c \rangle = \left(\begin{array}{cc}
\ 0 & 0 \\
\ d_c & 0
\end{array}\right), \quad \quad \langle \bar{\Delta}_c \rangle = \left(\begin{array}{cc}
\ 0 & \bar{d}_c \\
\ 0 & 0
\end{array}\right) $$
Now putting these vev's in the scalar potential Eq.(\ref{eq:Scalar-potential}) gives rise to
\begin{equation}
 V = m_l^2 (l_c^2)+M_2^2(\bar{d}_c^2)+f^2(l_c^4)+
4f^2 l_c^2 d_c^2 + v l_c^2 d_c^2 + f M l_c^2 \bar{d}_c + c.c
\label{eq:Scal-pot-vevs}
\end{equation}
Minimising the above potential the authors get the solution for
the parity breaking minima as
\begin{eqnarray}
 d_c &=& - \frac{v}{4 f^2}, \quad \bar{d}_c = \frac{f M l_c^2}{M_2^2} \nonumber \\
& & l_c^2 = \frac{(v^2 -4 f^2 m_l^2)M_2^2}{8f^4(M_2^2-M^2)}
\label{eq:PBS}
\end{eqnarray}
Since the original theory is Left-Right symmetric there exists an equivalent minima corresponding to $d_c \rightarrow d, \; \bar{d}_c \rightarrow \bar{d}, \; l_c \rightarrow l$. The degeneracy of these two equivalent vacua leads to the unavoidable consequences of formation
of domain walls. For a successful phase transition accompanying
the symmetry breaking, these domain walls should be unstable.
We follow the idea\cite{Rai:1992xw,Lew:1993yt} where it is argued that Planck sale suppressed non-renormalizable operators can potentially solve the domain wall
problem.

\subsection{Constraints on $M_R$ from domain wall removal}
We adopt the technique developed in\cite{Mishra:2009mk} to find the
operators suppressed by Planck scale. And we find the constrains
on the symmetry breaking scale from cosmological considerations.

We now find the $1/M_{Pl}$ terms in the effective potential by expanding
the K$\ddot{a}$hler potential and superpotential in powers of $1/M_{Pl}$. 
We include the terms containing $\Delta (\bar{\Delta}), \Delta_c 
(\bar{\Delta}_c), L (L_c)$ in the expansion. The K$\ddot{a}$hler 
potential in this model upto $1/M_{Pl}$ is 
\begin{eqnarray}
K &=& \text{Tr}[\Delta \Delta^{\dagger} +\bar{\Delta}
 \bar{\Delta}^{\dagger}]+\text{Tr}[\Delta_c \Delta^{\dagger}_c
 +\bar{\Delta}_c \bar{\Delta}^{\dagger}_c]+\frac{c_L}{M_{Pl}}(L^T\tau_2 
\Delta L +L^T\tau_2 \bar{\Delta}^{\dagger} L +h.c.) \nonumber \\
&& +\frac{c_R}{M_{Pl}}(L^T_c\tau_2 \Delta_c L_c +L^T_c\tau_2
\bar{\Delta}^{\dagger}_c L_c +h.c.) 
\end{eqnarray}

The superpotential upto the powers of $1/M_{Pl}$ is 
\begin{eqnarray}
\lefteqn{W = W_{ren} + \frac{a_L}{2M_{Pl}} (\text{Tr}[ \Delta 
\bar{\Delta}])^2 + \frac{a_R}{2M_{Pl}} (\text{Tr}[ \Delta_c 
\bar{\Delta}_c])^2 + \frac{b_L}{M_{Pl}} \text{Tr}[\Delta^2]\text{Tr}
[\bar{\Delta}^2]+\frac{b_R}{M_{Pl}} \text{Tr}[\Delta_c^2]
\text{Tr}[\bar{\Delta}_c^2]} \nonumber \\
&& +\frac{f_1}{M_{Pl}} (\text{Tr}[ \Delta \bar{\Delta}])
(\text{Tr}[ \Delta_c \bar{\Delta}_c])+\frac{f_2}{M_{Pl}}\text{Tr}
[\Delta^2]\text{Tr}[\Delta_c^2]+\frac{f_3}{M_{Pl}}\text{Tr}
[\bar{\Delta}^2]\text{Tr}[\bar{\Delta}_c^2]
+\frac{f_4}{M_{Pl}} (L^T L)(L_c^TL_c) \nonumber
\end{eqnarray}
Assuming a phase where only right type fields get non-zero vev and left type
fields get zero vev, the scalar potential upto the leading term in $1/M_{Pl}$ becomes,
\begin{equation}
 V^R_{eff} \sim \frac{2 f a_R }{M_{Pl}}l_c^2 \bar{d}_c^2 d_c +\frac{2a_Rm_{\Delta}}
{M_{Pl}} \bar{d}_c^3 d_c
\label{eq:V-R-eff}
\end{equation}
Using Eq.(\ref{eq:PBS}) in the above equation we get
\begin{equation}
  V^R_{eff} \sim -\frac{f a_R}{2M_{Pl}}\left(\frac{M^2}{M_2^4}+\frac{M^4}{M_2^6}\right)v\, l_c^6
\label{eq:V-R-eff-1}
\end{equation}

Similarly assuming non-zero vev for left type 
fields only and not for right type fields the effective potential becomes,
\begin{equation}
 V^L_{eff} \sim -\frac{f a_L}{2M_{Pl}}\left(\frac{M^2}{M_2^4}+\frac{M^4}{M_2^6}\right)v\, l^6
\end{equation}
If the scale of parity breaking is $M_R$ then $l_c = M_R$. In this case
where we consider the equal chance for left and right type fields getting
a vev, then $l=M_R$. So the effective energy difference arising from the
operators is given by,
\begin{equation}
 \delta \rho \sim \frac{f (a_L-a_R)}{2M_{Pl}}
\left( \frac{M^2}{M_2^4}+\frac{M^4}{M_2^6}\right)vM_R^6
\label{eq:eff-delta-rho}
\end{equation}
Now we shall compare this $\delta \rho$ with the case in a matter dominated era
where we have calculated the energy density for the domain wall to decay.
Before going further we make some approximations. The constants $v$ and $M$
in Eq.(\ref{eq:eff-delta-rho}) are the coefficients appearing in the 
trilinear terms in Eq.(\ref{eq:Scalar-potential}).  These trilinear 
terms are the soft terms. So the coefficients are of electroweak scale,
$M_{ew}$. $M_1^2$ and $M_2^2$ appear in the mass terms for $\Delta(\Delta_c)$ 
and $\bar{\Delta}(\bar{\Delta}_c)$. Since these triplet fields get their 
vev's at a scale $M_R$, the scales $M_1^2$ and $M_2^2$ should be of 
order $M_R$. But the scale of $M_R$ is higher than the electrowek scale, 
hence we find $M_2^2 > M^2$. So 
\begin{equation}
 \frac{M^2}{M_2^4} > \frac{M^4}{M_2^6}
\end{equation}
So the dominant term in Eq.(\ref{eq:eff-delta-rho}) is,
\begin{equation}
 \delta \rho \sim \frac{f (a_L-a_R)}{2M_{Pl}} \frac{M_{ew}^3}{M_R^4}M_R^6
\label{eq:dominat-delta-rho}
\end{equation}
Now by comparing,
\begin{equation}
 \frac{f (a_L-a_R)}{2M_{Pl}}M_{ew}^3 M_R^2 > M_R^4 \left( 
 \frac{M_R}{M_{Pl}}\right)^{3/2}
\end{equation}
Putting the electroweak scale as $M_{ew}\sim 10^3$GeV  and taking
$f$ as $O(1)$ we get the constraint on $(a_L-a_R)$ as
\begin{equation}
 (a_L-a_R) > 10^{-5}\left(\frac{M_R}{10^4{\rm GeV}}\right)^{7/2}
\label{eq:M_R-constrnt-MD}
\end{equation}
Comparing the obtained $\delta \rho$ with the case in a radiation
dominated era we get,
\begin{equation}
  \frac{f (a_L-a_R)}{2M_{Pl}}M_{ew}^3 M_R^2 > M_R^4 \left( 
 \frac{M_R}{M_{Pl}}\right)^2
\end{equation}
Proceeding as above we get the constraint on $(a_L-a_R)$ as
\begin{equation}
 (a_L-a_R) > 10^{-4} \left( \frac{M_R}{10^6 {\rm GeV}}\right)^4
\label{eq:M_R-constrnt-RD}
\end{equation}
Taking the dimensionless parameters $a_L, a_R$ to be of order one, 
the equation \ref{eq:M_R-constrnt-MD} gives an upper bound on the scale $M_R$ 
in a matter dominated era
\begin{equation}
M_R < 2.7 \times 10^{5} \text{GeV}
\label{MDbound}
\end{equation}
Similarly during the radiation dominated era, the equation \ref{eq:M_R-constrnt-RD}
gives an upper bound on $M_R$
\begin{equation}
M_R < 1 \times 10^7 \text{GeV}
\label{RDbound}
\end{equation}
Allowing further fine tuning between $a_L$ and $a_R$ will make this bound even more strict. However as we will see below, such a low intermediate D-parity breaking scale is 
not favored from successful gauge coupling unification point of view which makes this model less attractive.

Comparing the obtained $\delta \rho$ with the weak inflation case we have 
\begin{equation}
\frac{f (a_L-a_R)}{2M_{Pl}}M_{ew}^3 M_R^2 \geq  M^4_R \frac{T^{12}_D M^3_{Pl}}{M^{15}_R}
\end{equation}
Taking the dimensionless coefficients to be of order one, we arrive at the following bound on $M_R$
\begin{equation}
M_R \geq  1.4 \times 10^{5} T^{12/13}_D 
\end{equation}
Thus for $T_D$ of the order of electroweak scale, $M_R$ remains just below the gravitino bound. However, if $T_D > 1.44 \times \sim 10^4 \; \text{GeV}$, then the $M_R$ is forced to be higher 
that $10^9 \; \text{GeV}$ which, as noted in \cite{Mishra:2009mk} can be problematic if the reheating 
temperature after the domain wall disappearance is comparable to the temperature scale of original phase 
transition. In that case, the Universe would reheat to a temperature higher than $10^9 \; \text{GeV}$ giving 
rise to gravitino overabundance.

\subsection{Constraints on $M_R$ from Unification}
\begin{figure}
\begin{center}
 \includegraphics{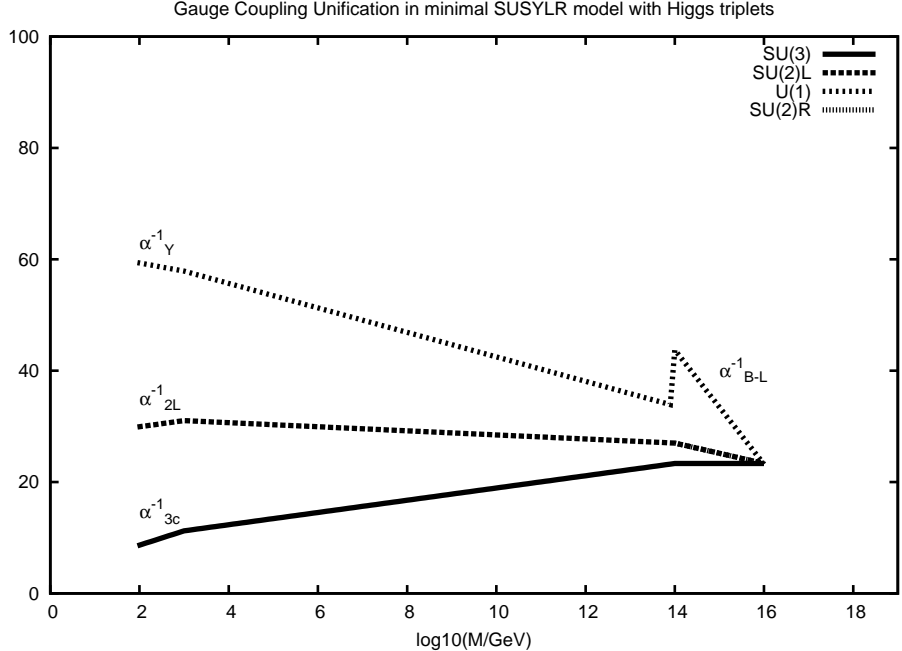}
\end{center}
\caption{Gauge coupling unification in minimal SUSYLR model with 
Higgs triplets, $M_{susy} = 1$ TeV, $M_R = 10^{14} $ GeV}
\label{fig1}
\end{figure}

Successful gauge coupling unification at scale $M_{G} > 10^{16} $ GeV puts
tight constraints on the intermediate symmetry breaking scales. Assuming the
Higgs triplets to be as heavy as the the scale $M_R$, we get a lower bound on
the scale $M_R$ to be of the order of $10^{14}$ GeV. A lower value of $M_R$ 
will make the couplings of $U(1)_{B-L}$ and $SU(2)_{L,R}$ meet before the allowed
unification scale from proton decay constraints. Although for the minimal 
particle content, the $SU(3)_c$ couplings do not meet the other two at one
point, we can always take into account of some additional fields in a Grand
Unified Theory (GUT) like $SO(10)$ which survive the GUT symmetry breaking
and can be as light as the scale $M_R$. Here we consider three pairs of extra 
heavy colored superfields 
$\chi(3,1,1,-\frac{2}{3}), \bar{\chi}(\bar{3},1,1,\frac{2}{3})$ which can 
be naturally fitted within $SO(10)$ GUT theory inside the 
representations $\textbf{120}, \overline{\textbf{126}}$. The resulting unification is shown in fig. \ref{fig1}. Thus the lower limit on $M_R$ from unification is in conflict with the bounds from domain wall removal (\ref{MDbound}),(\ref{RDbound}).

\section{Minimal Supersymmetric Left-Right Model with Higgs doublets}
\label{sec:SUSYLR2}
Spontaneous R-parity breaking can be achieved even without 
giving vev to the sneutrino fields. If the $U(1)_{B-L}$ symmetry is broken
by a Higgs field which has odd $B-L$ charge then R-parity is spontaneously broken. We call this model as Minimal Higgs Doublet (MHD) Model. The minimal such model \cite{Borah:2009ra,Borah:2011zz} has the following particle content
\begin{equation}
L(2,1,-1), \quad L_c(1,2,1), \quad S (1,1,0), \quad
Q(2,1,\frac{1}{3}),\quad  Q_c(1,2, -\frac{1}{3}) \nonumber
\end {equation}
\begin{equation}
H=
\left(\begin{array}{cc}
\ H^+_{L} \\
\ H^0_{L}/{\surd 2}
\end{array}\right)
\sim (2,1,1), \quad
H_c=
\left(\begin{array}{cc}
\ H^+_{R} \\
\ H^0_{R}/{\surd 2}
\end{array}\right)
\sim (1,2,-1), \nonumber
\end{equation}
\begin{equation}
\bar{H}=
\left(\begin{array}{cc}
\ h^0_{L}/{\surd 2} \\
\ h^-_{L}
\end{array}\right)
\sim (2,1,-1), \quad
\bar{H}_c=
\left(\begin{array}{cc}
\ h^0_{R}/{\surd 2} \\
\ H^-_{R}
\end{array}\right)
\sim (1,2,1), \nonumber
\end{equation}
\begin{equation}
\Phi_1(2,2,0), \quad \Phi_2(2,2,0) \nonumber 
\end{equation}
where the numbers in brackets correspond to the quantum numbers corresponding 
to $ SU(2)_L\times SU(2)_R \times U(1)_{B-L} $. The symmetry breaking pattern 
is 
\begin{eqnarray}
SU(2)_L \times SU(2)_R \times U(1)_{B-L} \quad
\underrightarrow{\langle H,H_c \rangle} \quad SU(2)_L \times U(1)_{Y} \quad
\underrightarrow{\langle \Phi \rangle}\quad U(1)_{em}
\end{eqnarray}
The renormalizable superpotential relevant for the spontaneous 
parity violation is given as follows
\begin{eqnarray}
\lefteqn{W_{ren}=h^{(i)}_l L^T\tau_2 \Phi_i \tau_2 L_c+ h^{(i)}_q Q^T\tau_2 
\Phi_i \tau_2 Q_c} \nonumber \\
&& +\mu_{ij}\text{Tr}\tau_2\Phi^T_i\tau_2\Phi_j + 
f_1(H^T\Phi_i H_c+\bar{H}^T\Phi_i \bar{H}_c)+ m_h H^T 
\tau_2 \bar{H} +m_h H^T_c \tau_2 \bar{H}_c
\end{eqnarray}
The scalar potential is $V = V_F+V_D+V_{soft}$ where 
$V_F = \lvert F_i \rvert^2, F_i = -\frac{\partial W}{\partial \phi} $ 
is the F-term scalar potential, $V_D = D^a D^a/2, D^a = 
-g(\phi^*_i T^a_{ij} \phi_j)$ is the D-term 
of the scalar potential and $ V_{soft}$ is the soft 
supersymmetry breaking scalar potential. We introduce the 
soft SUSY breaking terms to check if they alter relations between 
various mass scales in the model. The soft SUSY breaking 
superpotential in this case is given by
\begin{eqnarray}
\lefteqn{V_{soft} = m^2_{H} H^{\dagger} H+m^2_{H} \bar{H}^{\dagger} 
\bar{H}+m^2_{H} H^{\dagger}_c H_c+m^2_{H} \bar{H}^{\dagger}_c \bar{H}_c+m^2_{11}
 \Phi^{\dagger}_1 \Phi_1} \nonumber \\
&& +m^2_{22} \Phi^{\dagger}_2 \Phi_2+ (B_1H^{\dagger} \tau_2 
\bar{H} +B_2 H^{\dagger}_c \tau_2 \bar{H}_c + B \mu_{ij}\text{Tr}
[\tau_2 \Phi_i \tau_2 \Phi_j ]+h.c.) \nonumber \\
&& +(A_1 H^{\dagger} \Phi_i H_c +A_2 \bar{H}^{\dagger} 
\Phi_i \bar{H}_c +h.c.)
\label{soft}
\end{eqnarray}
where all the parameters $m_H, m_{11}, m_{22}, B, A$ are of the 
order of SUSY breaking scale $M_{susy} \sim \text{TeV}$.
We denote the vev of the neutral components of $\Phi_1, 
\Phi_2, H_L, \bar{H}_L, H_R, \bar{H}_R$ as $\langle (\Phi_1)_{11} 
\rangle = v_1,~ \langle (\Phi_2)_{22} \rangle = v_2,~ \langle H_L,\bar{H}_L \rangle 
= v_L,~ \langle H_R,\bar{H}_R \rangle =v_R$
Minimizing the potential with respect to $v_L, v_R$, we get the relations
\begin{eqnarray}
\lefteqn{\frac{\partial V}{\partial v_L}=\frac{1}{2}
(v_L(4v^2_R+v^2_1+v^2_2)f^2_1+2v_Rf_1(m_h(v_1+v_2)+2v_1(\mu_{11}+\mu_{12}) } \nonumber \\
&& 2v_2(\mu_{12}+\mu_{22}))+4m^2_Hv_L-2v_Lm^2_h+2v_LB_1+A_1v_1v_R) = 0
\label{eq1}
\end{eqnarray}
\begin{eqnarray}
\lefteqn{\frac{\partial V}{\partial v_R}=\frac{1}{2}
(v_R(4v^2_L+v^2_1+v^2_2)f^2_1+2v_Lf_1(m_h(v_1+v_2)+
2v_1(\mu_{11}+\mu_{12}) } \nonumber \\
&& 2v_2(\mu_{12}+\mu_{22}))+4m^2_Hv_R-2v_Rm^2_h+
2v_LB_2+A_1v_1v_L) = 0
\label{eq2}
\end{eqnarray}
From the above two equations we arrive at
\begin{eqnarray}
\lefteqn{v_R \frac{\partial V}{\partial v_L}-
v_L\frac{\partial V}{\partial v_R} = \frac{(v^2_R-v^2_L)}
{2}(4v_Lv_Rf^2_1+2f_1(m_h(v_1+v_2) } \nonumber \\
&& 2v_1(\mu_{11}+\mu_{12})+2v_2(\mu_{22}+\mu_{12}))+A_1v_1) = 0
\end{eqnarray}
Assuming $m_h, v_R \gg v_1, v_2, \mu_{ij} \sim M_{EW}$ the above
 relation gives the parity breaking solution ($v_L \neq v_R$)
$$ v_L \sim -\frac{m_h(v_1+v_2)}{2v_Rf_1} $$
From the above relations we can show that parity is broken 
spontaneously to give rise to the parity violating standard model. 
Also there is a seesaw between $v_L$ and $v_R$ from the above 
equation which can give rise to tiny neutrino masses \cite{Borah:2011zz}.
\begin{figure}
\begin{center}
\includegraphics{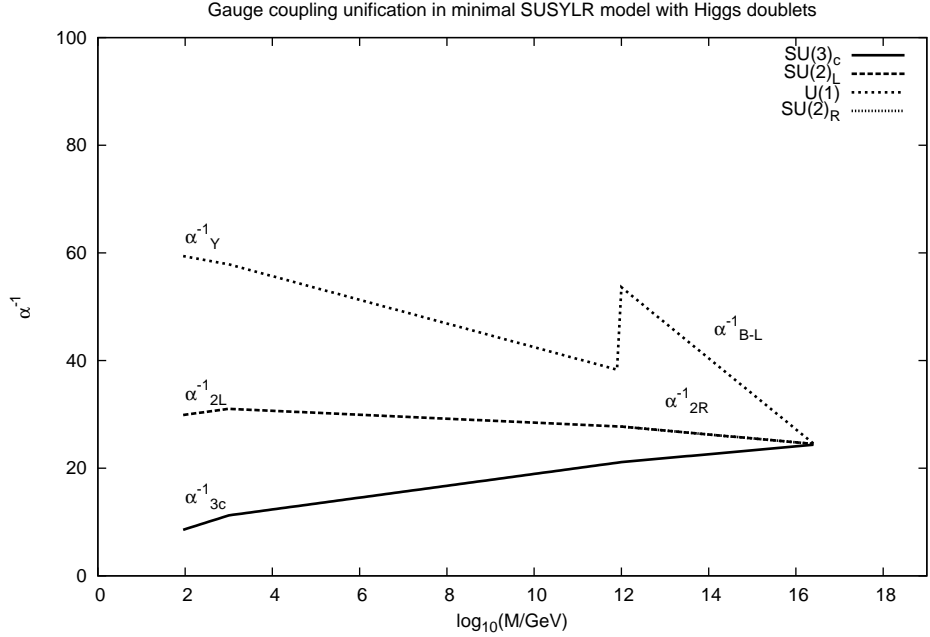}
\end{center}
\caption{Gauge coupling unification in minimal SUSYLR model with Higgs doublets, $M_{susy} = 1$ TeV, 
$M_R = 10^{12}$ GeV, $M_{GUT} = 10^{16.4}$ GeV}
\label{fig2}
\end{figure}
\subsection{Constraints on $M_R$ from domain wall removal}
\label{subsec:DW-constrnt-2}
Similar to the previous section, here also we find the $1/M_{Pl}$ terms 
in the effective potential by expanding the K$\ddot{a}$hler potential and 
superpotential in powers of $1/M_{Pl}$. The superpotential upto the powers 
of $1/M_{Pl}$ is 
\begin{eqnarray}
W &=& W_{ren} + \frac{a_L}{2M_{Pl}} (H^T \tau_2 \bar{H})^2+
\frac{a_R}{2M_{Pl}} (H^T_c \tau_2 \bar{H}_c)^2+\frac{b_L}{M_{Pl}}(H^TH)
(\bar{H}^2\bar{H})+\frac{b_R}{M_{Pl}}(H^T_cH_c)(\bar{H}^2_c\bar{H}_c) \nonumber \\
&& + \frac{f_2}{M_{Pl}} (H^T \tau_2 \bar{H})(H^T_c \tau_2 \bar{H}_c)+
\frac{f_3}{M_{Pl}} (H^TH)(H^T_cH_c)+\frac{f_4}{M_{Pl}} (\bar{H}^T\bar{H})
(\bar{H}^T_c \bar{H}_c) 
\end{eqnarray}
The K$\ddot{a}$hler potential in this model upto $1/M_{Pl}$ is 
$$ K = H^{\dagger}H +\bar{H}^{\dagger}\bar{H}+H^{\dagger}_c H_c +
\bar{H}^{\dagger}_c \bar{H}_c $$
Assuming a phase where only right type fields get non-zero vev 
and left type fields get zero vev, the scalar potential upto the 
leading term in $1/M_{Pl}$ becomes
\begin{equation}
 V^R_{eff}\sim (a_R +2b_R) m_h\frac{v^4_R}{M_{Pl}}
\end{equation}
Similarly for the phase where only left type fields get non-zero vev
\begin{equation}
  V^L_{eff} \sim (a_L +2b_L) m_h\frac{v^4_L}{M_{Pl}}
\end{equation}
Taking $m_h \sim M_R$ the effective energy difference can now be calculated as
\begin{equation}
 \delta \rho \sim [(a_R +2b_R) - (a_L +2b_L)]\frac{M_R^5}{M_{Pl}}
\end{equation}
Thus for the matter dominated era we have 
\begin{equation}
[(a_R +2b_R) - (a_L +2b_L)] > \frac{M^{1/2}_R}{M^{1/2}_{Pl}}
\end{equation}
And for the radiation dominated era
\begin{equation}
[(a_R +2b_R) - (a_L +2b_L)] > \frac{M_R}{M_{Pl}}
\end{equation}
Taking the various dimensionless parameters to be of order one, we
get the same upper bound on $M_R$ in both the above cases
\begin{equation}
M_R < M_{Pl} = 1 \times 10^{19} \text{GeV}
\end{equation}

Similarly, in the weak inflation scenario we have
\begin{equation}
[(a_R +2b_R) - (a_L +2b_L)]\frac{M_R^5}{M_{Pl}} \geq M^4_R \frac{T^{12}_D M^3_{Pl}}{M^{15}_R}
\end{equation}
Assuming the dimensionless coefficients to be order one, this leads to
\begin{equation}
M_R \geq 5.6 \times 10^{4} T^{3/4}_D
\end{equation}
Here $T_D$ can be as high as $5 \times 10^5 \; \text{GeV}$ for $M_R$ to remain below the gravitino bound. As noted in the
 previous section, $M_R > 10^9 \; \text{GeV}$ can lead to gravitino overabundance if the reheat temperature after 
the wall disappearance is same as the temperature of the original phase transition.

Thus the scale $M_R$ is less restrictive in this model compared to the 
SUSYLR model with Higgs triplets. As we will see below, one can have successful 
gauge coupling unification in this model for $M_R \geq  10^{12}$ GeV. Due to 
the possibility of successful removal of domain walls as well as successful 
gauge coupling unification, this model is the preferred one over the model with Higgs triplets.

\subsection{Constraints on $M_R$ from Unification}
\label{subsec:M-R-consrnt-2}
Similar to the minimal SUSYLR model with Higgs triplets, here also 
the intermediate symmetry breaking scales will be constrained by 
demanding successful gauge coupling unification at a very high
 scale $M_G ( >10^{16} \text{GeV})$. Similar to the previous case, 
here also the couplings of $U(1)_{B-L}$ and $SU(2)_{L,R}$ meet much 
before the allowed Unification scale if the intermediate symmetry 
breaking scale $M_R$ is lower than a certain value. For the minimal
 SUSYLR model with Higgs doublets, this lower bound on $M_R$ is found 
to be of the order of $10^{12}$ GeV. We also consider two additional 
heavy colored superfields so that the $SU(3)_c$ coupling meet the other 
two couplings at one point. They are denoted as $\chi(3,1,1,-\frac{2}{3}), 
\bar{\chi}(\bar{3},1,1,\frac{2}{3})$ and can be accommodated within $SO(10)$ 
GUT theory in the representations $\textbf{120},\overline{\textbf{126}}$. Here 
we assume that the structure of the GUT theory is such that these fields 
survive the symmetry breaking and can be as light as the $SU(2)_R$ breaking scale. 
The resulting gauge coupling unification as shown in the figure \ref{fig2}.

\section{Results and Conclusion}
\label{sec:con}
We have discussed the issue of domain wall formation due to the spontaneous breaking 
of D-parity in two different versions of 
supersymmetric left right models: one with Higgs triplets having $B-L$ charge $\pm 2$
and the other with Higgs doublets having $B-L$ charge $\pm 1$. Since stable domain walls 
are in conflict with cosmology, we consider the effects of Planck scale suppressed operators 
in destabilizing them. We consider the evolution and decay of domain walls in two different
epochs: radiation dominated as well as matter dominated. We find that successful removal of 
domain walls put rather strict constraints on the D-parity breaking scale $M_R$ in the model
with Higgs triplets. The model with Higgs doublets is far less
restrictive on the other hand. 
\begin{center}
\begin{table}[ht]
\caption{Bounds on $M_R/\text{GeV}$ in R-parity violating SUSYLR models}
\begin{tabular}{|c|c|c|c|c|}
\hline
Model  & Gauge Coupling  & DW removal & DW removal & DW removal \\
      &   Unification      & during MD era    & during RD era  & including WI\\
\hline
MSLRM & $ \geq 10^{14} $ & $  < 2.7 \times 10^{5}$ & $< 10^7$ & $\geq  1.4 \times 10^{5} T^{12/13}_D$ \\
MHD & $\geq 10^{12}$ & $< M_{Pl} $ & $< M_{Pl} $ & $\geq 5.6 \times 10^{4} T^{3/4}_D$ \\
BDM & $ \geq 1.5 \times 10^3$ & None & None & None  \\
\hline
\end{tabular}
\label{table1}
\end{table}
\end{center} 
\begin{center}
\begin{table}[ht]
\caption{Bounds on $M_R/\text{GeV}$ in R-parity conserving SUSYLR models}
\begin{tabular}{|c|c|c|c|c|}
\hline
Model  & Gauge Coupling  & DW removal & DW removal & DW removal \\
      & Unification      & during MD era    & during RD era  & including WI\\
\hline
ABMRS & $\geq 10^{15-16}$   &  $< 10^7$  & $< 10^{11}$    & $ \geq 8.6 \times 10^{4} T^{4/5}_D$  \\
BM & $ \geq 10^{14} $ &  None & None & None  \\
Bitriplet & $\geq 5 \times 10^{12} $ & None & None & None \\
\hline
\end{tabular}
\label{table2}
\end{table}
\end{center}
We also find the constraint on the D-parity breaking scale in these models by demanding 
gauge coupling unification at a scale $M_G > 10^{16}$ GeV. We use one-loop beta functions 
for both the models and take into account of some heavy colored superfields to make 
the $SU(3)_c$ coupling meet the other two exactly at one point. The results are shown in table (\ref{table1}). We also mention the model by Bhupal Dev and Mohapatra (BDM) \cite{Dev:2009aw}  in the table where the scale of $SU(2)_R \times U(1)_{B-L}$ symmetry breaking to $U(1)_Y$ denoted by $M_R$ can be as low as few TeV. However there is no constraint on the scale $M_R$ from domain wall disappearance due to the existence of a parity odd singlet which, after acquiring a vev breaks the degeneracy between two possible vacua. 

In table (\ref{table2}) we summarize the results of similar analysis obtained for R-parity conserving SUSYLR models in some of our earlier works. The bounds from domain wall disappearance in Aulakh-Bajc-Melfo-Rasin-Senjanovic (ABMRS) \cite{Aulakh:1998nn, Aulakh:1997ba} model and Babu-Mohapatra (BM) \cite{Babu:2008ep} model were discussed in \cite{Mishra:2009mk}. The bitriplet model \cite{Patra:2009wc} does not suffer from domain wall problem due to the existence of a parity odd singlet as pointed out in the introduction. The bounds on $M_R$ from gauge coupling unification in such models were discussed in \cite{Borah:2010kk,Borah:2010zq}.  

To summarize the result of this paper, it is shown that both domain wall 
removal and unification constraints can be satisfied in the Minimal Higgs doublet model or the MHD model whereas it is not 
possible to have successful removal of domain walls and gauge coupling unification together in the model 
with Higgs triplets (MSLRM).

\section{Acknowledgement}
We would like to thank Prof Urjit A. Yajnik, IIT Bombay for useful comments and discussions.

%

\bibliographystyle{apsrev}

\begin{thebibliography}{35}
\expandafter\ifx\csname natexlab\endcsname\relax\def\natexlab#1{#1}\fi
\expandafter\ifx\csname bibnamefont\endcsname\relax
  \def\bibnamefont#1{#1}\fi
\expandafter\ifx\csname bibfnamefont\endcsname\relax
  \def\bibfnamefont#1{#1}\fi
\expandafter\ifx\csname citenamefont\endcsname\relax
  \def\citenamefont#1{#1}\fi
\expandafter\ifx\csname url\endcsname\relax
  \def\url#1{\texttt{#1}}\fi
\expandafter\ifx\csname urlprefix\endcsname\relax\def\urlprefix{URL }\fi
\providecommand{\bibinfo}[2]{#2}
\providecommand{\eprint}[2][]{\url{#2}}

\bibitem[{\citenamefont{Kuchimanchi and Mohapatra}(1993)}]{Kuchimanchi:1993jg}
\bibinfo{author}{\bibfnamefont{R.}~\bibnamefont{Kuchimanchi}} \bibnamefont{and}
  \bibinfo{author}{\bibfnamefont{R.~N.} \bibnamefont{Mohapatra}},
  \bibinfo{journal}{Phys. Rev.} \textbf{\bibinfo{volume}{D48}},
  \bibinfo{pages}{4352} (\bibinfo{year}{1993}), \eprint{hep-ph/9306290}.

\bibitem[{\citenamefont{Pati and Salam}(1974)}]{Pati:1974yy}
\bibinfo{author}{\bibfnamefont{J.~C.} \bibnamefont{Pati}} \bibnamefont{and}
  \bibinfo{author}{\bibfnamefont{A.}~\bibnamefont{Salam}},
  \bibinfo{journal}{Phys. Rev.} \textbf{\bibinfo{volume}{D10}},
  \bibinfo{pages}{275} (\bibinfo{year}{1974}).

\bibitem[{\citenamefont{Mohapatra and Pati}(1975)}]{Mohapatra:1974gc}
\bibinfo{author}{\bibfnamefont{R.~N.} \bibnamefont{Mohapatra}}
  \bibnamefont{and} \bibinfo{author}{\bibfnamefont{J.~C.} \bibnamefont{Pati}},
  \bibinfo{journal}{Phys. Rev.} \textbf{\bibinfo{volume}{D11}},
  \bibinfo{pages}{2558} (\bibinfo{year}{1975}).

\bibitem[{\citenamefont{Senjanovic and Mohapatra}(1975)}]{Senjanovic:1975rk}
\bibinfo{author}{\bibfnamefont{G.}~\bibnamefont{Senjanovic}} \bibnamefont{and}
  \bibinfo{author}{\bibfnamefont{R.~N.} \bibnamefont{Mohapatra}},
  \bibinfo{journal}{Phys. Rev.} \textbf{\bibinfo{volume}{D12}},
  \bibinfo{pages}{1502} (\bibinfo{year}{1975}).

\bibitem[{\citenamefont{Mohapatra and Marshak}(1980)}]{Mohapatra:1980qe}
\bibinfo{author}{\bibfnamefont{R.~N.} \bibnamefont{Mohapatra}}
  \bibnamefont{and} \bibinfo{author}{\bibfnamefont{R.~E.}
  \bibnamefont{Marshak}}, \bibinfo{journal}{Phys. Rev. Lett.}
  \textbf{\bibinfo{volume}{44}}, \bibinfo{pages}{1316} (\bibinfo{year}{1980}).

\bibitem[{\citenamefont{Deshpande et~al.}(1991)\citenamefont{Deshpande, Gunion,
  Kayser, and Olness}}]{Deshpande:1990ip}
\bibinfo{author}{\bibfnamefont{N.~G.} \bibnamefont{Deshpande}},
  \bibinfo{author}{\bibfnamefont{J.~F.} \bibnamefont{Gunion}},
  \bibinfo{author}{\bibfnamefont{B.}~\bibnamefont{Kayser}}, \bibnamefont{and}
  \bibinfo{author}{\bibfnamefont{F.~I.} \bibnamefont{Olness}},
  \bibinfo{journal}{Phys. Rev.} \textbf{\bibinfo{volume}{D44}},
  \bibinfo{pages}{837} (\bibinfo{year}{1991}).

\bibitem[{\citenamefont{Fukuda et~al.}(2001)}]{Fukuda:2001nk}
\bibinfo{author}{\bibfnamefont{S.}~\bibnamefont{Fukuda}} \bibnamefont{et~al.}
  (\bibinfo{collaboration}{Super-Kamiokande}), \bibinfo{journal}{Phys. Rev.
  Lett.} \textbf{\bibinfo{volume}{86}}, \bibinfo{pages}{5656}
  (\bibinfo{year}{2001}), \eprint{hep-ex/0103033}.

\bibitem[{\citenamefont{Ahmad et~al.}(2002{\natexlab{a}})}]{Ahmad:2002jz}
\bibinfo{author}{\bibfnamefont{Q.~R.} \bibnamefont{Ahmad}} \bibnamefont{et~al.}
  (\bibinfo{collaboration}{SNO}), \bibinfo{journal}{Phys. Rev. Lett.}
  \textbf{\bibinfo{volume}{89}}, \bibinfo{pages}{011301}
  (\bibinfo{year}{2002}{\natexlab{a}}), \eprint{nucl-ex/0204008}.

\bibitem[{\citenamefont{Ahmad et~al.}(2002{\natexlab{b}})}]{Ahmad:2002ka}
\bibinfo{author}{\bibfnamefont{Q.~R.} \bibnamefont{Ahmad}} \bibnamefont{et~al.}
  (\bibinfo{collaboration}{SNO}), \bibinfo{journal}{Phys. Rev. Lett.}
  \textbf{\bibinfo{volume}{89}}, \bibinfo{pages}{011302}
  (\bibinfo{year}{2002}{\natexlab{b}}), \eprint{nucl-ex/0204009}.

\bibitem[{\citenamefont{Bahcall and Pena-Garay}(2004)}]{Bahcall:2004mz}
\bibinfo{author}{\bibfnamefont{J.~N.} \bibnamefont{Bahcall}} \bibnamefont{and}
  \bibinfo{author}{\bibfnamefont{C.}~\bibnamefont{Pena-Garay}},
  \bibinfo{journal}{New J. Phys.} \textbf{\bibinfo{volume}{6}},
  \bibinfo{pages}{63} (\bibinfo{year}{2004}), \eprint{hep-ph/0404061}.

\bibitem[{\citenamefont{Aulakh et~al.}(1998)\citenamefont{Aulakh, Melfo, and
  Senjanovic}}]{Aulakh:1998nn}
\bibinfo{author}{\bibfnamefont{C.~S.} \bibnamefont{Aulakh}},
  \bibinfo{author}{\bibfnamefont{A.}~\bibnamefont{Melfo}}, \bibnamefont{and}
  \bibinfo{author}{\bibfnamefont{G.}~\bibnamefont{Senjanovic}},
  \bibinfo{journal}{Phys. Rev.} \textbf{\bibinfo{volume}{D57}},
  \bibinfo{pages}{4174} (\bibinfo{year}{1998}), \eprint{hep-ph/9707256}.

\bibitem[{\citenamefont{Aulakh et~al.}(1997)\citenamefont{Aulakh, Benakli, and
  Senjanovic}}]{Aulakh:1997ba}
\bibinfo{author}{\bibfnamefont{C.~S.} \bibnamefont{Aulakh}},
  \bibinfo{author}{\bibfnamefont{K.}~\bibnamefont{Benakli}}, \bibnamefont{and}
  \bibinfo{author}{\bibfnamefont{G.}~\bibnamefont{Senjanovic}},
  \bibinfo{journal}{Phys. Rev. Lett.} \textbf{\bibinfo{volume}{79}},
  \bibinfo{pages}{2188} (\bibinfo{year}{1997}), \eprint{hep-ph/9703434}.

\bibitem[{\citenamefont{Babu and Mohapatra}(2008)}]{Babu:2008ep}
\bibinfo{author}{\bibfnamefont{K.~S.} \bibnamefont{Babu}} \bibnamefont{and}
  \bibinfo{author}{\bibfnamefont{R.~N.} \bibnamefont{Mohapatra}},
  \bibinfo{journal}{Phys. Lett.} \textbf{\bibinfo{volume}{B668}},
  \bibinfo{pages}{404} (\bibinfo{year}{2008}), \eprint{0807.0481}.

\bibitem[{\citenamefont{Patra et~al.}(2009)\citenamefont{Patra, Sarkar, Sarkar,
  and Yajnik}}]{Patra:2009wc}
\bibinfo{author}{\bibfnamefont{S.}~\bibnamefont{Patra}},
  \bibinfo{author}{\bibfnamefont{A.}~\bibnamefont{Sarkar}},
  \bibinfo{author}{\bibfnamefont{U.}~\bibnamefont{Sarkar}}, \bibnamefont{and}
  \bibinfo{author}{\bibfnamefont{U.}~\bibnamefont{Yajnik}},
  \bibinfo{journal}{Phys. Lett.} \textbf{\bibinfo{volume}{B679}},
  \bibinfo{pages}{386} (\bibinfo{year}{2009}), \eprint{0905.3220}.

\bibitem[{\citenamefont{Kibble}(1980)}]{Kibble:1980mv}
\bibinfo{author}{\bibfnamefont{T.~W.~B.} \bibnamefont{Kibble}},
  \bibinfo{journal}{Phys. Rept.} \textbf{\bibinfo{volume}{67}},
  \bibinfo{pages}{183} (\bibinfo{year}{1980}).

\bibitem[{\citenamefont{Hindmarsh and Kibble}(1995)}]{Hindmarsh:1994re}
\bibinfo{author}{\bibfnamefont{M.~B.} \bibnamefont{Hindmarsh}}
  \bibnamefont{and} \bibinfo{author}{\bibfnamefont{T.~W.~B.}
  \bibnamefont{Kibble}}, \bibinfo{journal}{Rept. Prog. Phys.}
  \textbf{\bibinfo{volume}{58}}, \bibinfo{pages}{477} (\bibinfo{year}{1995}),
  \eprint{hep-ph/9411342}.

\bibitem[{\citenamefont{Rai and Senjanovic}(1994)}]{Rai:1992xw}
\bibinfo{author}{\bibfnamefont{B.}~\bibnamefont{Rai}} \bibnamefont{and}
  \bibinfo{author}{\bibfnamefont{G.}~\bibnamefont{Senjanovic}},
  \bibinfo{journal}{Phys. Rev.} \textbf{\bibinfo{volume}{D49}},
  \bibinfo{pages}{2729} (\bibinfo{year}{1994}), \eprint{hep-ph/9301240}.

\bibitem[{\citenamefont{Lew and Riotto}(1993)}]{Lew:1993yt}
\bibinfo{author}{\bibfnamefont{H.}~\bibnamefont{Lew}} \bibnamefont{and}
  \bibinfo{author}{\bibfnamefont{A.}~\bibnamefont{Riotto}},
  \bibinfo{journal}{Phys. Lett.} \textbf{\bibinfo{volume}{B309}},
  \bibinfo{pages}{258} (\bibinfo{year}{1993}), \eprint{hep-ph/9304203}.

\bibitem[{\citenamefont{Mishra and Yajnik}(2010)}]{Mishra:2009mk}
\bibinfo{author}{\bibfnamefont{S.}~\bibnamefont{Mishra}} \bibnamefont{and}
  \bibinfo{author}{\bibfnamefont{U.~A.} \bibnamefont{Yajnik}},
  \bibinfo{journal}{Phys. Rev.} \textbf{\bibinfo{volume}{D81}},
  \bibinfo{pages}{045010} (\bibinfo{year}{2010}), \eprint{0911.1578}.

\bibitem[{\citenamefont{Chang et~al.}(1984{\natexlab{a}})\citenamefont{Chang,
  Mohapatra, and Parida}}]{Chang:1983fu}
\bibinfo{author}{\bibfnamefont{D.}~\bibnamefont{Chang}},
  \bibinfo{author}{\bibfnamefont{R.~N.} \bibnamefont{Mohapatra}},
  \bibnamefont{and} \bibinfo{author}{\bibfnamefont{M.~K.}
  \bibnamefont{Parida}}, \bibinfo{journal}{Phys. Rev. Lett.}
  \textbf{\bibinfo{volume}{52}}, \bibinfo{pages}{1072}
  (\bibinfo{year}{1984}{\natexlab{a}}).

\bibitem[{\citenamefont{Chang et~al.}(1984{\natexlab{b}})\citenamefont{Chang,
  Mohapatra, and Parida}}]{Chang:1984uy}
\bibinfo{author}{\bibfnamefont{D.}~\bibnamefont{Chang}},
  \bibinfo{author}{\bibfnamefont{R.~N.} \bibnamefont{Mohapatra}},
  \bibnamefont{and} \bibinfo{author}{\bibfnamefont{M.~K.}
  \bibnamefont{Parida}}, \bibinfo{journal}{Phys. Rev.}
  \textbf{\bibinfo{volume}{D30}}, \bibinfo{pages}{1052}
  (\bibinfo{year}{1984}{\natexlab{b}}).

\bibitem[{\citenamefont{Sarkar et~al.}(2008)\citenamefont{Sarkar, Abhishek, and
  Yajnik}}]{Sarkar:2007er}
\bibinfo{author}{\bibfnamefont{A.}~\bibnamefont{Sarkar}},
  \bibinfo{author}{\bibnamefont{Abhishek}}, \bibnamefont{and}
  \bibinfo{author}{\bibfnamefont{U.~A.} \bibnamefont{Yajnik}},
  \bibinfo{journal}{Nucl. Phys.} \textbf{\bibinfo{volume}{B800}},
  \bibinfo{pages}{253} (\bibinfo{year}{2008}), \eprint{0710.5410}.

\bibitem[{\citenamefont{Dev and Mohapatra}(2010)}]{Dev:2009aw}
\bibinfo{author}{\bibfnamefont{P.~S.~B.} \bibnamefont{Dev}} \bibnamefont{and}
  \bibinfo{author}{\bibfnamefont{R.~N.} \bibnamefont{Mohapatra}},
  \bibinfo{journal}{Phys. Rev.} \textbf{\bibinfo{volume}{D81}},
  \bibinfo{pages}{013001} (\bibinfo{year}{2010}), \eprint{0910.3924}.

\bibitem[{\citenamefont{Borah et~al.}(2011)\citenamefont{Borah, Patra, and
  Sarkar}}]{Borah:2010zq}
\bibinfo{author}{\bibfnamefont{D.}~\bibnamefont{Borah}},
  \bibinfo{author}{\bibfnamefont{S.}~\bibnamefont{Patra}}, \bibnamefont{and}
  \bibinfo{author}{\bibfnamefont{U.}~\bibnamefont{Sarkar}},
  \bibinfo{journal}{Phys. Rev.} \textbf{\bibinfo{volume}{D83}},
  \bibinfo{pages}{035007} (\bibinfo{year}{2011}), \eprint{1006.2245}.

\bibitem[{\citenamefont{Borah}(2011)}]{Borah:2011zz}
\bibinfo{author}{\bibfnamefont{D.}~\bibnamefont{Borah}}, \bibinfo{journal}{Int.
  J. Mod. Phys.} \textbf{\bibinfo{volume}{A26}}, \bibinfo{pages}{1305}
  (\bibinfo{year}{2011}).

\bibitem[{\citenamefont{Vilenkin}(1985)}]{Vilenkin:1984ib}
\bibinfo{author}{\bibfnamefont{A.}~\bibnamefont{Vilenkin}},
  \bibinfo{journal}{Phys. Rept.} \textbf{\bibinfo{volume}{121}},
  \bibinfo{pages}{263} (\bibinfo{year}{1985}).

\bibitem[{\citenamefont{Kawasaki and Takahashi}(2005)}]{Kawasaki:2004rx}
\bibinfo{author}{\bibfnamefont{M.}~\bibnamefont{Kawasaki}} \bibnamefont{and}
  \bibinfo{author}{\bibfnamefont{F.}~\bibnamefont{Takahashi}},
  \bibinfo{journal}{Phys. Lett.} \textbf{\bibinfo{volume}{B618}},
  \bibinfo{pages}{1} (\bibinfo{year}{2005}), \eprint{hep-ph/0410158}.

\bibitem[{\citenamefont{Lyth and Stewart}(1995)}]{Lyth:1995hj}
\bibinfo{author}{\bibfnamefont{D.~H.} \bibnamefont{Lyth}} \bibnamefont{and}
  \bibinfo{author}{\bibfnamefont{E.~D.} \bibnamefont{Stewart}},
  \bibinfo{journal}{Phys. Rev. Lett.} \textbf{\bibinfo{volume}{75}},
  \bibinfo{pages}{201} (\bibinfo{year}{1995}), \eprint{hep-ph/9502417}.

\bibitem[{\citenamefont{Lyth and Stewart}(1996)}]{Lyth:1995ka}
\bibinfo{author}{\bibfnamefont{D.~H.} \bibnamefont{Lyth}} \bibnamefont{and}
  \bibinfo{author}{\bibfnamefont{E.~D.} \bibnamefont{Stewart}},
  \bibinfo{journal}{Phys. Rev.} \textbf{\bibinfo{volume}{D53}},
  \bibinfo{pages}{1784} (\bibinfo{year}{1996}), \eprint{hep-ph/9510204}.

\bibitem[{\citenamefont{Gonzalez-Garcia and Nir}(1989)}]{GonzalezGarcia:1989zh}
\bibinfo{author}{\bibfnamefont{M.~C.} \bibnamefont{Gonzalez-Garcia}}
  \bibnamefont{and} \bibinfo{author}{\bibfnamefont{Y.}~\bibnamefont{Nir}},
  \bibinfo{journal}{Phys. Lett.} \textbf{\bibinfo{volume}{B232}},
  \bibinfo{pages}{383} (\bibinfo{year}{1989}).

\bibitem[{\citenamefont{Romao and Nogueira}(1990)}]{Romao:1989yh}
\bibinfo{author}{\bibfnamefont{J.~C.} \bibnamefont{Romao}} \bibnamefont{and}
  \bibinfo{author}{\bibfnamefont{P.}~\bibnamefont{Nogueira}},
  \bibinfo{journal}{Phys. Lett.} \textbf{\bibinfo{volume}{B234}},
  \bibinfo{pages}{371} (\bibinfo{year}{1990}).

\bibitem[{\citenamefont{Amsler et~al.}(2008)}]{Amsler:2008zzb}
\bibinfo{author}{\bibfnamefont{C.}~\bibnamefont{Amsler}} \bibnamefont{et~al.}
  (\bibinfo{collaboration}{Particle Data Group}), \bibinfo{journal}{Phys.
  Lett.} \textbf{\bibinfo{volume}{B667}}, \bibinfo{pages}{1}
  (\bibinfo{year}{2008}).

\bibitem[{\citenamefont{Beall et~al.}(1982)\citenamefont{Beall, Bander, and
  Soni}}]{Beall:1981ze}
\bibinfo{author}{\bibfnamefont{G.}~\bibnamefont{Beall}},
  \bibinfo{author}{\bibfnamefont{M.}~\bibnamefont{Bander}}, \bibnamefont{and}
  \bibinfo{author}{\bibfnamefont{A.}~\bibnamefont{Soni}},
  \bibinfo{journal}{Phys. Rev. Lett.} \textbf{\bibinfo{volume}{48}},
  \bibinfo{pages}{848} (\bibinfo{year}{1982}).

\bibitem[{\citenamefont{Borah and Patra}(2009)}]{Borah:2009ra}
\bibinfo{author}{\bibfnamefont{D.}~\bibnamefont{Borah}} \bibnamefont{and}
  \bibinfo{author}{\bibfnamefont{S.}~\bibnamefont{Patra}}
  (\bibinfo{year}{2009}), \eprint{0910.0146}.

\bibitem[{\citenamefont{Borah and Yajnik}(2011)}]{Borah:2010kk}
\bibinfo{author}{\bibfnamefont{D.}~\bibnamefont{Borah}} \bibnamefont{and}
  \bibinfo{author}{\bibfnamefont{U.~A.} \bibnamefont{Yajnik}},
  \bibinfo{journal}{Phys. Rev.} \textbf{\bibinfo{volume}{D83}},
  \bibinfo{pages}{095004} (\bibinfo{year}{2011}), \eprint{1010.6289}.

\end{thebibliography}

\end{document}